\documentclass[dvips]{article}

\usepackage{icrc2011}

\title{Searching for Dark Matter Annihilation in M87}
\newcommand{\etal}{\MakeLowercase{\textit{et al. }}} % "et al."
\shorttitle{Saxena \etal Dark Matter Annihilation in M87}

\authors{Sheetal Saxena$^{1}$, Dominik Els\"{a}sser$^{1}$, Michael R\"{u}ger $^{1}$, Alexander Summa $^{1}$, Karl Mannheim $^{1}$ }
\afiliations{$^1$Institute for Theoretical Physics and Astrophysics, University of W\"{u}rzburg, Campus Hubland Nord, Emil-Fischer-Str. 31, 97074 W\"{u}rzburg, Germany\\  }
\email{saxena@astro.uni-wuerzburg.de}

\abstract{Clusters of galaxies, such as the Virgo cluster, host enormous quantities of dark matter, making them prime targets for efforts in indirect dark matter detection via potential radiative signatures from annihilation of dark matter particles and subsequent radiative losses of annihilation products. However, a careful study of ubiquitous astrophysical backgrounds is mandatory to single out potential evidence for dark matter annihilation. Here, we construct a multiwavelength spectral energy distribution for the central radio galaxy in the Virgo cluster, M87, using a state-of-the-art numerical Synchrotron Self Compton approach. Fitting recent Chandra, Fermi-LAT and Cherenkov observations, we probe different dark matter annihilation scenarios including a full treatment of the inverse-Compton losses from electrons and positrons produced in the annihilation. It is shown that such a template can substantially improve upon existing dark matter detection limits. }

\begin{document}
\maketitle

%Begin the section.
\section{Introduction}

The presence of dark matter in the universe can be astrophysically observed through the spectrum of acoustic resonances in the primordial plasma imprinted on the spectrum of anisotropies in the cosmic microwave background radiation, and its gravitational effects on visible matter. This dark matter could be weakly interacting massive particles (WIMPs) with masses at the electroweak symmetry breaking scale $m_{\chi}={\mathcal O}\left(\left(G_{F}\sqrt{2}\right)^{-\frac{1}{2}}\right)=246\,\mathrm{GeV}$, freezing-out in the early universe. Their relic density would be in agreement with the observed $\Omega_{dm}h^{2}=0.1120$ \cite{lab1}. This is termed the ``WIMP Miracle''. Moreover, stable neutral particles at the electroweak symmetry breaking scale are predicted by supersymmetric extensions of the Standard Model. \\
In the present day universe, annihilation interactions in regions with high mass densities still occur and can lead to the production of energetic neutrinos, electrons, positrons and gamma rays which inverse-Compton scatter with starlight photon fields or the cosmic microwave background resulting in signals potentially detectable across the soft to hard X-ray energy band. Limits on this dark matter annihilation (DMA) emission have been studied thus far in the Galactic Center \cite{lab2}, extragalactic gamma ray background \cite{lab9}, clusters of galaxies \cite{lab3} and by focusing on dwarf galaxies in the Milky Way halo \cite{lab4, lab5}. Searching for signatures of secondary gamma rays and decay products lends insight into the thermally averaged annihilation cross section and mass of the dark matter particle. \\
The first radio galaxy detected in the TeV energy range, M87 is one of the best studied in its source class with a jet resolved by optical, radio and X-ray observations. This cD galaxy is the nearest radio galaxy emitting very high energy (VHE) gamma rays and is known for its proximity to the Earth ($D = 16\,\mathrm{Mpc}$) and exceptionally bright arcsecond scale jet \cite{lab6}.  M87 is an excellent laboratory in which to study dark matter because of its location in the central high mass density region of the Virgo cluster. In this work, the imprint of WIMP annihilation radiation from the halo structure of M87 in the form of prompt and inverse-Compton gamma rays is investigated, and dark matter particles are constrained, via a multiwavelength analysis of the spectral energy distribution of M87.

\section{Observations and Analysis}

 \begin{figure*}[!t]
   \centerline{\includegraphics[width=1\linewidth]{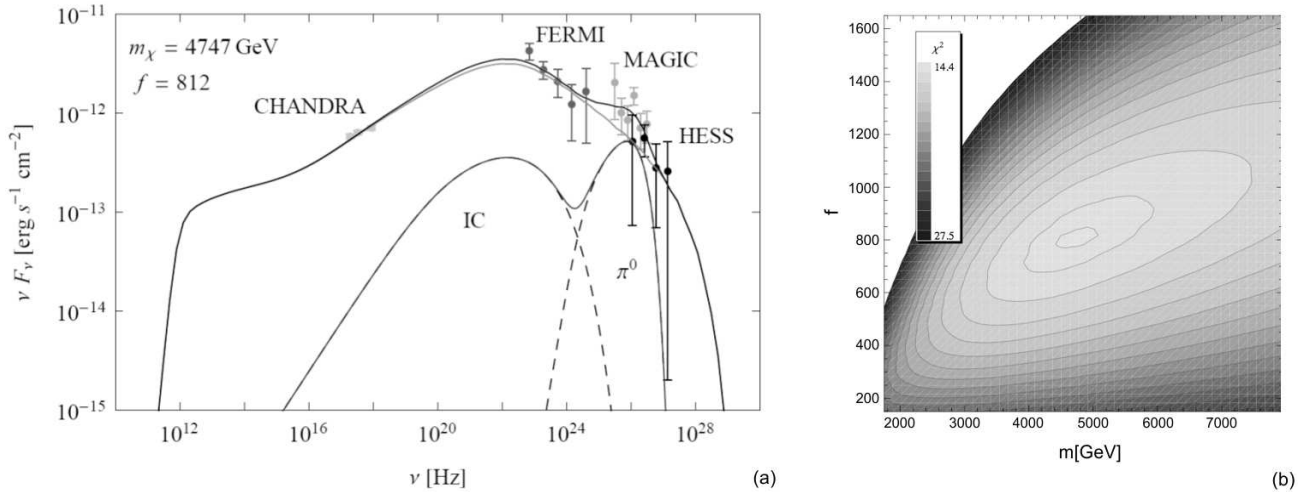}\label{fig1}}
   \caption{(a) Spectral energy distributions of M87 using the Synchrotron Self Compton model (light grey line); and with inclusion of dark matter annihilation model (dark line). The SSC model alone produces a $\chi^{2}$ value of 27.5 whereas an improvement to 14.4 is achieved using the dark matter annihilation model for a particle of mass 4.7 TeV and a boost factor of 812. The data points used in this analysis are shown, with inverse-Compton and prompt pion gamma peaks also illustrated. There is a noticeable gap between observed data and the SSC model in the MAGIC regime which is closed by the SSC+DMA model. (b) Island plot showing result of $\chi^{2}$ analysis for the combined M87 fit. The boost factor $f$ is introduced as a free parameter to account for the enhancement due to sub-halo clumping; $m$ denotes the generic WIMP mass.
            }
   \label{double_fig}
 \end{figure*}

Observations from the Chandra X-ray Observatory, Fermi Large Area Telescope (Fermi-LAT), Major Atmospheric Gamma-ray Imaging Cherenkov Telescope (MAGIC), and High Energy Stereoscopic System (H.E.S.S.) Cherenkov array are used in this analysis. In contrast to the model presented by Abdo et al. \cite{lab6}, historical data in the low-energy range from the radio-to-optical regime which can be contaminated by dust and starlight and are thought to originate from different regimes in the source than the VHE-emission are not included in the fit but rather used as benchmarks that the emission should not exceed. Although non-contemporaneous, the data set represents a long-term average by excluding significant flaring events and thus represents the steady low state spectral energy distribution of high-energy emission in M87. \\
The Synchrotron Self Compton (SSC) process occurs when electrons accelerated in the magnetized plasma jet emerging from the ergosphere of a central black hole reach ultra-relativistic speeds and emit synchrotron photons which then inverse-Compton scatter off the original source population of electrons. This generates a high-energy spectrum that extends into the TeV regime. The Synchrotron Self Compton model of R\"{u}ger et al. \cite{lab7} is applied here to the high-energy data with a focus near the upper cut-off of the spectrum so that the Compton scattering cross section is treated appropriately in the Klein-Nishina regime. This model has relativistic electrons injected into a spherical emission region with a randomly oriented magnetic field, moving up the jet with relativistic speed. Taking a bulk Lorentz factor $\Gamma=2.3$ (Doppler factor $\delta=3.9$) as in \cite{lab6} we vary other parameters so the maximum Lorentz factor $\Gamma_{\max}=10^{8}$, magnetic field $B=3\,\rm{G}$, normalization factor $K=10^{6}\,\rm{cm^{-3}s^{-1}}$, and the radius of the emitting source $R=3.5\times10^{13}\,\rm{cm}$ to obtain a differential slope of $s=2.2$ for the injected electron power law distribution with exponential cut-off. \\
A dark matter model is then added to this SSC model. When dark matter neutralinos annihilate, they produce heavy quarks, leptons and W bosons. During their subsequent hadronization, they decay mainly into pions. Prompt pion emission is the emittance of very high energy gamma rays from the decay of neutral pions. The charged pions decay into electrons and positrons which can up-scatter cosmic microwave background photons through the inverse-Compton mechanism also to very high energies. This model is referred to here as SSC+DMA. \\
The differential flux of gamma rays produced by the decay of annihilation products in M87
is given by Equation (1).  $\left\langle \sigma_{A}\nu\right\rangle$ is the thermally averaged annihilation cross section, $m_{\chi}$ the mass of the dark matter particle, and $f$ the boost factor that accounts for enhancement due to sub-halo clumping of dark matter in the M87 halo. $D$ is the distance from the Earth to M87, $\rho(r)$ the dark matter density profile and $\frac{\mathrm{d}N_{\gamma}}{\mathrm{d}E}$ the gamma photon spectrum coming from the decay of neutral pions from hadronization in the annihilation process (prompt pion emission). 

\begin{equation}
\left(\frac{\mathrm{d}\Phi}{\mathrm{d}E}\right)_{\pi^{0}}=\frac{1}{4\pi}\frac{f\left\langle\sigma_{A}\,\nu\right\rangle}{2m_{\chi}^{2}}\frac{\mathrm{d}N_{\gamma}}{\mathrm{d}E}\frac{1}{{D}^{2}}\int\limits_{\rm{M87}} \!  \mathrm{d}V\,\rho^{2}(r)
\end{equation}

The differential flux of high-energy photons produced by inverse-Compton scattering of charged particles off the cosmic microwave background is given by Equation (2). $P\left(E,E'\right)$ is the differential power emitted into photons of energy $E$ by an electron or positron with energy $E'$. $b\left(E'\right)$ is the total rate of electron/positron energy loss due to inverse-Compton scattering as in \cite{Cirelli}, and $\frac{\mathrm{d}N_{e}}{\mathrm{d}\widetilde{E}}$ is the spectrum of secondary electrons and positrons with energy $\widetilde{E}$. \\  \\

\begin{eqnarray}
\left(\frac{\mathrm{d}\Phi}{\mathrm{d}E}\right)_{\rm{IC}}&=& \frac{1}{E}\frac{f\left\langle\sigma_{A}\,\nu\right\rangle}{4\pi m_{\chi}^{2}}\frac{1}{{D}^{2}}\int\limits_{\rm{M87}} \!  \mathrm{d}V\,\rho^{2}\left(r\right) \times \nonumber \\
&&\int\limits_{m_{e}}^{m_{\chi}}\mathrm{d}E'\,\frac{P\left(E,E'\right)}{b\left(E'\right)}\int\limits_{E'}^{m_{\chi}}\mathrm{d}\widetilde{E}\,\frac{\mathrm{d}N_{e}}{\mathrm{d}\widetilde{E}}
\end{eqnarray}

Finally a Navarro-Frenk-White density profile is assumed for the dark matter halo so that $\rho(r)=\rho_{NFW}(r)$ \cite{lab8}, a generic species of annihilating WIMPs with rest mass in the GeV - TeV range, and a thermally averaged annihilation cross section $\left\langle \sigma_{A}\nu\right\rangle=3\times10^{-24}\,\rm{cm}^3\rm{s}^{-1}$. $10^{6}$ realizations of generic WIMP neutralinos motivated by Supersymmetry were generated to obtain the average secondary electron/positron spectrum $\frac{\mathrm{d}N_{e}}{\mathrm{d}\widetilde{E}}$. The gamma ray emission spectrum due to the decay of neutral pions in the annihilation process was generated numerically using the DarkSUSY code \cite{Gondolo}. The prompt pion emission and inverse-Compton mechanism reflect the dark matter halo distribution while usual Synchrotron Self Compton processes originate from the jet.

The best-fit combined model of AGN emission and the complete dark matter related emission is determined using a $\chi^{2}$-test. See Figure 1.

\section{Results}

In addition to closing the noticeable gap between the observational data and the SSC model in the high-energy regime, introducing a WIMP annihilation component to the Synchrotron Self Compton emission model of multi-wavelength observations of M87 produces a minimal $\chi^2$ value and thus an improved agreement with the data set. The best-fit employed a neutralino mass $m_{\chi}=4.7\,\rm{TeV}$ and boost factor $f=812$ both within the reach of numerical models and modern collider experiments.\\
Assuming annihilation interactions of WIMPS with TeV-scale masses in the dark matter halo of M87 are responsible for prompt gamma ray and inverse-Compton emission components, this method can be used to constrain viable dark matter models as new observations are being made eg. through Fermi-LAT and MAGIC telescope observations. Furthermore, insight into the discrimination between diffuse halo emission and jet emission as well as sub-halo clumping effects can be found through ongoing observational and theoretical investigation of the spatial distribution of the gamma ray emission in M87. The gamma ray luminosity due to dark matter annihilation in M87 can reach the same level as the emission from its relativistic jet for reasonable boost factors. \\
Since the high-energy emission of M87 shows blazar-like variability \cite{lab10}, the constraints for dark matter models can be improved significantly with future contemporaneous multiwavelength data, measuring the spectral energy distribution during low flux states. Whereas the gamma rays from the jet show correlated broad-band variability, the quiescent components produced by pions and inverse-Compton scattering electrons should remain steady. \emph{The DMA prompt pion and inverse-Compton emission result in a characteristic double-hump structure superimposed on the SSC emission which could in future be used as a distinguishing characteristic.} A dark matter induced component to the spectral energy distribution would stand out as steady emission. Thus there is potential to recognize this during low state emission of the active galactic nucleus and possibly extract dark matter properties such as the WIMP mass, annihilation cross section and boost factor due to substructure. Among the steady background components, the only other potentially confusing source of gamma rays is due to cosmic rays traveling through the intracluster medium. This component should, however, be well distinguishable owing to its smooth power law and different spatial profile \cite{lab11}. Although the significance of this result is not sufficient to claim evidence of a specific particle it is encouraging further studies, especially since previous deep exposures of other targets like the Galactic Center \cite{Horns} have also provided tentative hints of TeV-scale WIMP annihilation.

\clearpage

\end{document}